\documentstyle[12pt]{article}
\newlength{\dinwidth}
\newlength{\dinmargin}
\newlength{\extraspace}
\newlength{\extraspaces}
\newcommand{\be}{\begin{equation}
\addtolength{\abovedisplayskip}{\extraspaces}
\addtolength{\belowdisplayskip}{\extraspaces}
\addtolength{\abovedisplayshortskip}{\extraspace}
\addtolength{\belowdisplayshortskip}{\extraspace}}
\newcommand{\ee}{\end{equation}}
\newcommand{\bdm}{\begin{displaymath}
\addtolength{\abovedisplayskip}{\extraspaces}
\addtolength{\belowdisplayskip}{\extraspaces}
\addtolength{\abovedisplayshortskip}{\extraspace}
\addtolength{\belowdisplayshortskip}{\extraspace}}
\newcommand{\edm}{\end{displaymath}}
\renewcommand{\thefootnote}{\fnsymbol{footnote}}
\def\simlt{\mathrel{\lower2.5pt\vbox{\lineskip=0pt\baselineskip=0pt
           \hbox{$<$}\hbox{$\sim$}}}}
\newcommand{\beq}{\begin{equation}}
\newcommand{\eeq}{\end{equation}}
\newcommand{\bea}{\begin{eqnarray}}
\newcommand{\eea}{\end{eqnarray}}
\newcommand{\ts}{\thinspace}

\newcommand{\pr}{Phys.\ Rev.\ }

\newcommand{\pl}{Phys.\ Lett.\ {\bf B}}

\newcommand{\nnbar}{\nu_\tau{\overline \nu}_\tau}

\newcommand{\tautaub}{\tau{\overline \tau}}
\newcommand{\gev}{\ts \rm GeV \ts}

\newcommand{\gfiv}{\gamma^5}

\newcommand{\gmu}{\gamma^{\mu}}
\begin{document}
\begin{titlepage}
\begin{flushright}
UTPT-95-15\\
hep-ph/9509272\\
%\Large{DRAFT}
\end{flushright}
\vspace{24mm}
\begin{center}
\Large{{\bf  New flavor physics in $b$ decays}}
\end{center}
\vspace{5mm}
\begin{center}
B. Holdom\footnote{e-mail address: holdom@utcc.utoronto.ca} and
M.~V.~Ramana\footnote{e-mail address:
ramana@medb.physics.utoronto.ca}\\ {\normalsize\it Department of
Physics}\\ {\normalsize\it University of Toronto}\\ {\normalsize\it
Toronto,
Ontario,}\\ {\normalsize CANADA, M5S 1A7}
\end{center}
\vspace{2cm}
\thispagestyle{empty}
\begin{abstract}

A new $U(1)$ gauge boson coupling predominantly to the third family has
been considered in connection with recent LEP data.  We consider another
likely consequence of such a gauge boson, a greatly enhanced $b$ quark
decay mode, $b \to s \nnbar$.

\end{abstract}
\end{titlepage}
\newpage

\renewcommand{\thefootnote}{\arabic{footnote}}
\setcounter{footnote}{0}
\setcounter{page}{2}

Broken family gauge symmetries often arise in theories which seek to
explain the masses of fermions in a dynamical framework.  The lightest
of these gauge bosons are expected to couple to the third family and
not to the lighter families.  Previously \cite{hint,vertex}, it was
shown that the mixing of a such a massive $U(1)$ gauge boson (the $X$
boson) with the $Z$ causes shifts in the $Z$ couplings to the third
family.  This leads to a distinctive pattern of universality-breaking
corrections which is quite consistent with precision electroweak
measurements. In particular the presently observed anomaly in
${R}_{b}$ and the discrepency between the values of ${\alpha }_{s}$
determined from $R_\ell$ and low energy measurements were shown to be
natural consequences.

The $X$ boson with coupling strength $g_X$ receives a mass from the same
source as the $W$ and $Z$ masses, and this immediately leads \cite{hint}
to the relation\footnote{More conservatively this could be taken as an
upper bound on the coupling to mass ratio, since in principle there could
be additional contributions to the $X$ boson mass.}
\beq
\label{xzrelation}
{\left({{\frac{{g}_{X}}{{M}_{X}}}}\right)}^{2}={\frac{G}{2\sqrt {2}}}.
\eeq
Results follow
without requiring knowledge of $M_X$, although we may imagine $M_X$ to
lie in the several hundred GeV to 1 TeV range.

When the $Z$ couplings to the third family are shifted by new flavor
physics, then flavor changing vertices such as $Zb{\overline s}$,
$Zd{\overline s}$ and $Zb{\overline d}$ may be generated by the
effects of fermion mass mixing.  The resulting flavor changing neutral
currents from $Z$ exchange were considered briefly in \cite{hint}, and
in more detail in \cite{zhang}.  Although very dependent on the nature
of the fermion mass mixing, some of these effects may lie close to
experimental limits.

In the present work we are instead considering
the effect of a tree-level $X$-boson exchange, with the vertex
$Xb{\overline s}$ generated via fermion mass mixing.  This effect is not
suppressed by the $Z$-$X$ mixing amplitude, and would lead
to the decays $b \to s \nnbar$ and $b \to s \tautaub$.
These decays need only compete against the standard model $b \to c$
decay which is already suppressed by a small $V_{cb}$.  Given the poor
experimental limits on FCNC's involving the heavier quarks alone, it
is even possible that these supposedly rare decay modes could occur at
rates approaching the standard semileptonic $b$ decay rate.

Indeed, there is some suggestion of a discrepancy
between the observed experimental value of the semileptonic branching
ratio of the $B$ meson \cite{pdata} and theoretical predictions of the
same \cite{bigi}. Related to this discrepancy are the indications that the
charm multiplicity of the decay products is low \cite{roudeau}.
Whether these discrepancies are a sign of new physics or are just a
function of our poor understanding of the hadronic physics involved is
still subject to debate \cite{kagan,bagan}.  It should be noted that the
$Z$-$X$ mixing effect makes a low $\alpha_s(M_Z)$ in the 0.11 range more
likely \cite{hint}, and this in turn exacerbates the discrepancy in the
semileptonic branching ratio \cite{shifman}.

In \cite{hint} it was argued that the $X$-boson should couple to the
following current ($R_\mu,L_\mu \equiv \gamma_\mu(1\pm{\gamma
}_{5})/2$)\footnote{In \cite{hint,vertex} a reversed, nonstandard
definition was assumed.} \beq {J}_{\mu }^{X}=\overline{t}
(L_\mu-R_\mu)t+\overline{b}(L_\mu-R_\mu)b+\overline{\tau
}(L_\mu+R_\mu)\tau +{\overline{\nu }}_{\tau}L_\mu{\nu }_{\tau}\eeq with
coupling strength $g_X$.  Non-zero CKM
matrix elements imply that there is mixing between quarks of
different families, and we shall assume that some of this mixing happens
in the down quark sector.  This can be expressed in terms of the mixing
matrices $L^d$ and $R^d$ which act on the mass eigenstate bases
$(d,s,b)_L$ and $(d,s,b)_R$ respectively (such that the CKM matrix is
$L^{u\dagger}L^d$).
Then the $X$ boson couplings to quarks contains the following $b-s$
transitions, \beq g_X
X^{\mu}(\lambda^L_{23}{\overline s}L_\mu b - \lambda^R_{23}{\overline
s}R_\mu b + {\rm h.c.}),\eeq
where $\lambda^L_{ij} \equiv L_{3i}^{d*}L_{3j}^d$ and $\lambda^R_{ij}
\equiv R_{3i}^{d*}R_{3j}^d$.

It is worth noting that there are no significant experimental
constraints on FCNC's involving just the second and third generation
quarks.  This is in contrast to the very strong constraints on FCNC's
involving the $d$ quark.  We shall henceforth assume that the 13 and 12
components of $\lambda^L$ and $\lambda^R$ are
very small as required; of more interest to us are the much less
constrained 23 components.

The interesting decays mediated by the $X$ boson are $b
\to s \nnbar$ and $b \to s \tautaub$. The decay $b \to s \tautaub$
is suppressed by phase space compared to the decay $b \to s \nnbar$;
we estimate the
ratio of the $\nnbar$ rate to the $\tautaub$ rate to be about $4.5$
for $m_b = 4.5 \gev$ and about $2.5$ for $m_b = 4.8 \gev$. Furthermore
reconstruction of the $\tautaub$ decay mode is challenging, making a clear
experimental signature difficult.  A possible exception is when both
$\tau$'s decay leptonically, leading to $B\rightarrow
{X}_{s}{\ell}^{+}{\ell}^{-}+$ (missing energy) with $\ell = e,\mu$.

In the following we will concentrate on the $b \to s \nnbar$ mode.
In the standard model this mode proceeds via penguin diagrams
with a virtual $Z$ boson and box diagrams involving $W$ bosons.  The
predicted branching ratio for this process is about $5
\times 10^{-5}$ \cite{ali,babar}. Presently there appear to be no
significant experimental bounds \cite{pdata}.

The $X$ boson induces the following effective four-fermion interaction.
\beq (g_X/M_X)^2
(\lambda^L_{23}{\overline s}L_\mu b - \lambda^R_{23}{\overline s}R_\mu b)
{\overline{\nu }}_{\tau}L^\mu{\nu }_{\tau}\eeq
The contribution of this term to the decay rate is
\beq
\Gamma(b \to s \nnbar) = (|\lambda^L_{23}|^2+|\lambda^R_{23}|^2) G^2
m_b^5/{12288\pi^3}.
\label{nuwidth}
\eeq
Eq.~(\ref{xzrelation}) has been used and $m_s$ has been ignored.

We may compare to the semileptonic width of the $b$ and include QCD
corrections to obtain
\beq
{{\Gamma(b \to s \nnbar)}\over{\Gamma(b \to c e \nu)}} =
{{|\lambda^L_{23}|^2+|\lambda^R_{23}|^2} \over
{64|V_{cb}|^2g({{m_c^2}\over{m_b^2}})}}\left[
{\frac{1-{\frac{2{\alpha }_{s}({m}_{b})}{3\pi }}f(0)}{1-{\frac{2{\alpha
}_{s}({m}_{b})}{3\pi }}f({\frac{{m}_{c}}{{m}_{b}}})}}\right]
\label{ratio}
\eeq
where $f(x)$ and $g(x) = 1-8x+8x^3-x^4-12x^2{\rm ln}(x)$ are found in
\cite{cm}.
For example, with $m_b = 4.8 \gev$ and $m_c = 1.5 \gev$,
$|\lambda^L_{23}|^2+|\lambda^R_{23}|^2\approx
30 {\left|{{V}_{cb}}\right|}^2$ would give a $b \to s \nnbar$ branching
ratio equal to the
semileptonic branching ratio of $b$ quarks. Mixing angles of this size in
the down quark sector may be somewhat larger than expected, but are
not entirely unreasonable.  If they were this large they would be
sufficient to account for the possible discrepancy in the semileptonic
branching ratio.  This is because the new decay mode would increase the
total width of the $b$ such as to decrease the semileptonic branching
ratio by 10 or 15 percent.

We consider $B\overline{B}$ production in ${e}^{+}{e}^{-}$ collisions,
with one of the $B$ mesons decaying through $b \to s \nnbar$.  This
strange quark should hadronize into an
energetic $K$ or $K^*$ a large fraction of the time. One then observes an
energetic strange meson, a number of other particles from the decay of
the second $B$ meson and a large amount of missing energy and
momentum.  We note that since there
are two neutrinos in the event, the quantity $E_{miss}^2 - |\vec
p_{miss}|^2$ should not peak near zero, unlike the case
of only one neutrino. After identifying the energetic strange meson, the
second $B$ meson may be reconstructed from its decay products if it decays
purely hadronically.  For this one may
require that there be no charged lepton in the event.  A veto on events
where the energetic strange meson arises from a displaced vertex due to a
charm meson could also be useful. In the case that the two $B$ mesons are
produced close to rest, such as at CLEO, the energetic $K$ or $K^*$ would
recoil against the total missing momentum of the event.  In the other case
when the $B$ mesons are energetically produced, the $K$ or $K^*$ will be
clearly isolated from the decay products of the second $B$.

The other main effect of nonvanishing $\lambda^L_{23}$ and
$\lambda^R_{23}$ occurs in $B_s -
{\overline B_s}$ mixing. The main contribution of the $X$ boson comes from
the operator
\beq
{{(\lambda^L_{23}+\lambda^R_{23})^2}\over
{8}}{{g_X^2}\over{M_X^2}}{\overline s}\gmu \gfiv b{\overline s}\gmu \gfiv
b  + {\rm h.c.},
\eeq
which gives a mass mixing of order
\beq
\Delta M_{B_s} \approx G f_{B_s}^2
M_{B_s}|\lambda^L_{23}+\lambda^R_{23}|^2/8\sqrt{2}.
\eeq
The only experimental result comes from time dependence of $B_s -
{\overline B_s}$ mixing using dileptons, which yields $\Delta M_{B_s}
> 1.2 \times 10^{-12} \gev$ \cite{pdata}.  The standard model result
is larger, $\Delta M_{B_s} = (1 \pm .5) \times 10^{-10}\gev$
\cite{ali2}, while the $X$ boson contribution with $
|\lambda^L_{23}+\lambda^R_{23}| = 2|V_{cb}| $ is larger still, $\Delta
M_{B_s} \sim
10^{-9} \gev$. This provides additional incentive to try to get some
experimental handle on $B_s - {\overline B_s}$ mixing.  We note that a
similar argument using the observed size of $B_d - {\overline B_d}$ mixing
produces the very tight constraint $|\lambda^L_{13}
+\lambda^R_{13}|<0.002$ (with a weaker constraint on $|\lambda^L_{13}
-\lambda^R_{13}|$).

If there is mixing in the charged lepton sector, then besides $b \to s
\tau^-\tau^+$ there could also be $X$ boson contributions to $b \to
s\tau^-l^+$ and $b \to s l^-l^+$ ($l = e,\mu$).  These contributions
are naturally small since the vertices involved bring in additional
small mixing angles.  For example if mixing in the lepton sector was
similar to the mixing in the quark sector then a
suppression of order
${\left|{{V}_{cb}}\right|}^{4}$ for $b \to s l^-l^+$ relative to
$b \to s \nnbar$ would not be surprising.  The upper limits from
CLEO\cite{karen} are $BR(B \to K^* e^+ e^-) < 1.9 \times 10^{-5}$,
$BR(B \to K^* \mu^+ \mu^-) < 3.9\times 10^{-5}$ and $BR(B
\to K^* e^+ \mu^-) < 1.8\times 10^{-5}$. Similarly CDF quotes an upper
limit of $BR(B \to K^* \mu^+ \mu^-) < 5.1\times 10^{-5}$ \cite{carol}.
The standard model predictions for these are of order $5 \times
10^{-6}$ \cite{ali}.

In conclusion we have given some motivation for imagining that certain
``rare'' decay modes of the $b$ quark may in fact be very substantial
decay modes.  Our speculations tie in closely with the hints of new
flavor physics presently emerging from LEP.

{\bf Acknowledgements}

M.V.R. would like to thank M.~Luke and U.~Amaldi for useful
discussions as well as the Boston University Physics Department for
its hospitality while some of this work was done.  This research was
supported in part by the Natural Sciences and Engineering Research
Council of Canada.

%\newpage

\end{document}